\documentclass{PoS}
\usepackage{lineno}

\newcommand{\pt}{$p_T$}
\newcommand\Sl{\tilde l}
\newcommand{\sq}{$\sqrt{s}$}

\newcommand{\fbinv}{fb$^{-1}$}

\usepackage[symbol]{footmisc}

\title{Searches for Dark Matter at the LHC in forward proton mode}

\ShortTitle{DM searches in the forward proton mode}

\author{\speaker{Marek Ta\v{s}evsk\'{y}}\\
        Institute of Physics of the Czech Academy of Sciences,
        Na Slovance 2, Prague, 18221, Czech Republic\\
        E-mail: \email{Marek.Tasevsky@cern.ch}}

\author{L.A. Harland-Lang\\
  Rudolf Peierls Centre, Beecroft Building, Parks Road, Oxford, OX1 3PU,
  UK\\
E-mail: \email{lucian.harland-lang@physics.ox.ac.uk}}

\author{V. A. Khoze\\
  IPPP, Department of Physics, University of Durham, Durham, DH1 3LE, UK\\
  Petersburg Nuclear Physics Institute, NRC ``Kurchatov Institute'', Gatchina, St.~Petersburg, 188300, Russia\\
  E-mail: \email{V.A.Khoze@durham.ac.uk}}

\author{M. G. Ryskin\\
        Petersburg Nuclear Physics Institute, NRC ``Kurchatov Institute'', Gatchina, St.~Petersburg, 188300, Russia\\
        E-mail: \email{ryskin@thd.pnpi.spb.ru}}
  
\abstract{We analyze in detail the LHC prospects at the center-of-mass energy of
  14~TeV for charged electroweakino searches, decaying to leptons, in compressed
  supersymmetry scenarios, via exclusive photon-initiated pair production. This
  provides a potentially increased sensitivity in comparison to inclusive
  channels, where the background is often overwhelming. We pay particular
  attention to the challenges that such searches would face in the hostile high
  pile--up environment of the LHC, giving close consideration to the backgrounds
  that will be present. The signal we focus on is the exclusive production of
  same-flavour muon and electron pairs, with missing energy in the final state,
  and with two outgoing intact protons registered by the dedicated forward
  proton detectors installed in association with ATLAS and CMS. We present
  results for slepton masses of 120--300 GeV and slepton-neutralino mass
  splitting of 10-20 GeV, and find that the relevant backgrounds can be
  controlled to the level of the expected signal yields. The most significant
  such backgrounds are due to semi-exclusive lepton pair production at lower
  masses, with a proton produced in the initial proton dissociation system
  registering in the forward detectors, and from the coincidence of forward
  protons produced in pile-up events with an inclusive central event that mimics
  the signal. We also outline a range of potential methods to further suppress
  these backgrounds as well as to enlarge the signal yields.}

\FullConference{
European Physical Society Conference on High Energy Physics - EPS-HEP2019 -\\
                        10-17 July, 2019\\
                        Ghent, Belgium}

\begin{document}
%\linenumbers
\section{Introduction}
This text describes only the main points of the analysis which has been
published in Ref.~\cite{Harland-Lang:2018hmi} (for details and references, see
this paper).

One of the main goals of the physics program at the LHC and future colliders is
the search for  beyond the Standard Model (BSM) physics. A possibility that has
received significant recent attention in the context of the LHC and future collider
searches is the electroweak pair production of $R$-parity conserving states in
compressed mass scenarios of supersymmetry (SUSY). That is, where the mass
difference between the heavier state (e.g. the chargino, $\widetilde{\chi}^\pm$,
or slepton, $\tilde l(g)$) and the Lightest SUSY Particle (LSP)
$\widetilde{\chi}^0_{1}$ is small, see references in Ref.~\cite{Harland-Lang:2018hmi}.
%for instance \cite{LeCompte:2011cn,Baer:2014yta,Moortgat-Picka:2015yla,Khachatryan:2016mbu,CMS:2016zvj,Golling:2016gvc,Lehtinen:2017vdt,Khoze:2017ixx,Aaboud:2017leg,Sirunyan:2018lul,Sirunyan:2017lae,Sirunyan:2018iwl,Bagnaschi:2017tru}.
Searches in the compressed mass region via standard inclusive channels are
experimentally very challenging, in particular because the SM $WW$ background
produces a very similar final state and the visible decay products have low
momenta and therefore often do not pass detector acceptance thresholds. In order
to trigger on such events, generally the presence of an additional jet or photon
due to initial state radiation is required, providing the final-state particles
with a boost in the transverse plane, and thus generating a large missing
transverse momentum.

The potential to search for these comparatively light charged SUSY particles via
photon--initiated production in hadron collisions has been widely discussed over
the past few decades~\cite{Ohnemus:1993qw,Piotrzkowski:2000rx,Schul:2008sr,Khoze:2010ba,HarlandLang:2011ih}. One clear benefit is its model independence, in
sharp contrast to many other reactions. That is, the production cross sections
are directly predicted in terms of the electric charges of the relevant states.
Now, the development of the forward proton detectors (FPD) at the LHC allows us
to perform a wide program of such searches~\cite{Albrow:2008pn,N.Cartiglia:2015gve}.
%Kepka:2008yx,Royon:2016hdx,Harland-Lang:2016kog,Khoze:2017igg,Baldenegro:2018hng,Harland-Lang:2015cta,Harland-Lang:2018iur}.
In particular, dedicated AFP~\cite{AFP,Tasevsky:2015xya} and PPS~\cite{CT-PPS} FPDs have recently been installed in association with both ATLAS and CMS,
respectively. The purpose of these near-beam detectors is to measure intact
protons arising at small angles, giving access to a wide range of Central
Exclusive Production (CEP) processes
\begin{equation}
pp\to p~+~X~+~p\ ,
\end{equation}
where the plus sign indicates the presence of the Large Rapidity Gaps (LRG)
between the produced state and outgoing protons. The experimental signature for
the CEP of electroweakinos is then the presence of two very forward protons that
are detected in the FPD and two leptons from the slepton
$\tilde l(g)\to l+\widetilde{\chi}^0$ decay whose  
production vertex is indistinguishable from the primary vertex measured in the
central detector. The well--defined initial state and presence of the tagged
outgoing protons provides a unique handle, completely absent in the inclusive
case, that is able to greatly increase the discovery potential.

In Ref.~\cite{Harland-Lang:2018hmi} we studied in detail the LHC prospects for searching for such exclusive slepton pair production in compressed mass scenarios at the center-of-mass energy of \sq\ = 14~TeV.
%Such a possibility was first discussed in~\cite{Khoze:2017igg}  (see also~\cite{vak,vak1,vak2,Marektalk}), and has more recently been considered in~\cite{Beresford:2018pbt}.
We performed for the first time a systematic analysis of the various
challenges and sources of backgrounds that such studies must deal with, a
serious consideration of which is essential to assess the potential of these
exclusive channels. In particular, as well as the irreducible exclusive $WW$
background, we also consider the reducible backgrounds from semi--exclusive
lepton pair production, where a proton produced in the initial proton
dissociation registers in the FPDs, and the pile--up background where two soft
inelastic events coincide with an inelastic lepton pair production event.
%As we
%will see these two formally reducible backgrounds are expected to play a
%significant role during nominal LHC running conditions, and require close
%examination.

%\begin{figure}
%  \includegraphics[width=0.3\textwidth]{fig_01a.pdf}\hspace*{0.6cm}
%  \includegraphics[width=0.3\textwidth]{fig_01b.pdf}\hspace*{0.6cm}
%  \includegraphics[width=0.3\textwidth]{fig_01c.pdf}  
%  \caption{Schematic illustrations of (left) single-diffractive dissociation
%    (SD), (middle) double-diffractive dissociation (DD) and (right) central
%    diffraction (CD) and the kinematic variables used to describe them.}
%  \label{sigbg}
%\end{figure}

\section{Signal cross section and selection cuts}
We consider the direct pair production of smuons and selectrons
$\Sl_{L,R}$ ($l=e,\mu$) only, where the subscripts $L, R$ denote the left- and
right-handed partner of the electron or muon. The four sleptons are assumed to
be mass degenerate and to decay with a 100\% branching ratio into the
corresponding SM partner lepton and $\widetilde{\chi}^0_{1}$ neutralino. We take four slepton
mass points, 120, 200, 250 and 300 GeV, with in each case a mass splitting of
$\Delta M = M_{\Sl} - M_{\widetilde{\chi}^0_1}=10$~GeV and 20~GeV.
For all exclusive processes below, we use the \textsc{SuperChic} 2.07 Monte
Carlo (MC) generator~\cite{Harland-Lang:2015cta}. All applied cuts in this
analysis are summarized in table~\ref{cuts} and can be divided into three
classes, namely: cuts on the
detected protons in the FPDs (`FPD cuts'); requiring no other additional charged
particles in the central detector (`no-charged cuts'); and the selection applied
to the lepton pair (`di--leptons cuts').

\begin{table}
\begin{tabular}{||c|c|c||} 
\hline
&$5<p_{T,l_1,l_2} <$ 40~GeV &$|\eta_{l_1,l_2}| < 2.5$ (4.0) \\ \cline{2-3}
&Aco $\equiv 1-|\Delta\phi_{l_1l_2}|/\pi >$ 0.13 (0.095) &$2 < m_{l_ll_2} < 40$~GeV \\ \cline{2-3}
Di--lepton&$\Delta R (l_1,l_2) > 0.3$ & $|\eta_{l_1}-\eta_{l_2}| < 2.3$ \\ \cline{2-3}
&$\bar{\eta}\equiv |\eta_{l_1}+\eta_{l_2}|/2 < 1.0$&$||\vec{p_{Tl_1}}|-|\vec{p_{Tl_2}}|| > 1.5$~GeV\\ \cline{2-3}
& $W_{\rm miss} > 200$~GeV & \\ \hline\hline
FPD& $0.02 < \xi_{1,2} < 0.15$ &$p_{T,{\rm proton}} < 0.35$~GeV \\ \hline\hline
No--charge& No hadronic activity & z-veto \\ \hline
\end{tabular}
\caption{\small{Cuts used in this analysis.}}
\label{cuts}
\end{table}

\section{Backgrounds}
The predicted signal production cross sections are small, being $\sim 0.1$~fb or
less after accounting for all relevant cuts and efficiencies, depending on the
slepton mass in the experimentally allowed region. Therefore it is essential to
collect these events at nominal LHC luminosities and for any backgrounds to be
under very good control. We have considered: the irreducible photon--initiated
$WW$ background; the reducible background from the semi--exclusive
photon--initiated production of lepton pairs and QCD--initiated production of
gluon and $c$--quark jets (via leptonic decays of hadrons produced in
hadronization) at low mass, where a proton produced in the initial proton
dissociation registers in the forward proton detectors; the reducible pile--up
background where (dominantly) two independent single--diffractive events
coincide with an inelastic lepton pair production event. For the proton
dissociation and pile--up backgrounds we have performed dedicated MC
simulations, including most of relevant detector effects and efficiencies, in
order to evaluate their impact as accurately as possible. To evaluate the effect
of pile--up backgrounds we generate the dominant source of inclusive lepton
production, due to non--diffractive (ND) jet production, with both
\textsc{Pythia}~8.2~\cite{Sjostrand:2014zea} and \textsc{Herwig}~7.1~\cite{Bahr:2008pv,Bellm:2015jjp}.

We have found that requiring that the lepton pair lie in the signal $m_{ll}<40$
GeV region, combined with further judicially chosen cuts on the lepton momenta
leads to significant reductions in the background. The pile--up backgrounds are
strongly reduced by the use of time-of-flight (ToF) subdetectors in the FPDs
(cuts to suppress pile--up backgrounds in general are elaborated in Ref~\cite{Tasevsky:2014cpa,Cox:2007sw}), as well as the
aforementioned lepton cuts and a further cut on the proton transverse momentum.
These also help to reduce the semi--exclusive backgrounds considerably.
%However,
%after accounting for all of these effects we find that the backgrounds from
%pile--up and semi--exclusive photon--initiated lepton pair production are
%nonetheless expected to be of the same order as the signal, with the irreducible
%$WW$ background being somewhat lower. 
\vspace*{-0.2cm}
\section{Results}\label{FY}
\begin{table}[h]
\begin{center}
\begin{tabular}{|l||c|c|c||} 
  \hline
Event yields / & \multicolumn{3}{c||}{$\langle \mu \rangle_{PU}$} \\ 
\cline{2-4} 
$\cal L = $ 300~\fbinv\ & 0 & 10  & 50 \\ \hline\hline
Excl. sleptons & 0.6---2.9 & 0.5---2.4 & 0.3---1.4 \\ \hline
Excl. $l^+l^-$  & 1.4 & 1.2 & 0.7 \\ \hline
Excl. $K^+K^-$  & $\sim 0$ & $\sim 0$ & $\sim 0$ \\ \hline
Excl. $W^+W^-$  & 0.7 & 0.6 & 0.3 \\ \hline
Excl. $c\bar{c}$& $\sim 0$ & $\sim 0$ & $\sim 0$ \\ \hline
Excl. $gg$      & $\sim 0$ & $\sim 0$ & $\sim 0$ \\ \hline
Incl. ND jets & $\sim 0$/$\sim 0$ & 0.1/0.1 & 1.8/2.4  \\ \hline
\end{tabular}
\caption{\small{The event yields corresponding to an integrated luminosity of
    300~\fbinv\ as a function of pile--up amount for the slepton signal and all
    considered backgrounds. 
  All numbers correspond to the di--lepton mass range
  $2 < m_{ll} < 40$~GeV and lepton \pt\ $>$ 5~GeV and a tracker coverage of
  $|\eta| <2.5$. The ranges
  in the signal event yields illustrate the
  spread obtained from the entire studied slepton mass range: the lower value
  comes from the $M_{\Sl} = 300$~GeV, the higher from the $M_{\Sl} = 120$~GeV
  scenario.}}
\label{AllEY2.5}
\end{center}
\end{table}
We collect our results for the expected signal and background event yields in
tables~\ref{AllEY2.5} and~\ref{AllEY4}. Here, the former case corresponds to
$|\eta| < 2.5$ (i.e. the current tracker coverage) while the latter corresponds
to $|\eta| < 4.0$ (i.e. the tracker coverage for the Run III upgrade of ATLAS
\cite{Collaboration:2017mtb} and CMS \cite{Klein:2017nke}). To give a global
picture, these results correspond to the full di--lepton mass range of 
$2 < m_{ll} < 40$~GeV, although information about individual lepton \pt\ ranges
for processes where it is relevant can be found in tables~3, 4 and 5 of
Ref~\cite{Harland-Lang:2018hmi}. 
In summary, we observe that in total 2--3 signal events for 300~\fbinv  \,can be
expected, with a $S/B \sim 1$.
We note that \textsc{Pythia} 8.2 and \textsc{Herwig} 7.1 give similar
predictions for the contamination from the inclusive ND jets. 
%MTThese relatively small numbers therefore clearly do not correspond to a statistically significant observation.
% $\sim$ 3.0~$\sigma$. 
 %The ways to increase it are, however, numerous.
There are however various ways to improve the $S/B$ ratio.
\begin{table}
\begin{center}
\begin{tabular}{|l||c|c|c||} 
  \hline
Event yields / & \multicolumn{3}{c||}{$\langle \mu \rangle_{PU}$} \\ 
\cline{2-4} 
$\cal L = $ 300~\fbinv\ & 0 & 10  & 50 \\ \hline\hline
Excl. sleptons & 0.6---3.0 & 0.5---2.6 & 0.3---1.5 \\ \hline
Excl. $l^+l^-$  & 1.1 & 0.9 & 0.5 \\ \hline
Excl. $K^+K^-$  & $\sim 0$ & $\sim 0$ & $\sim 0$ \\ \hline
Excl. $W^+W^-$  & 0.6 & 0.5 & 0.3 \\ \hline
Excl. $c\bar{c}$& $\sim 0$ & $\sim 0$ & $\sim 0$ \\ \hline
Excl. $gg$      & $\sim 0$ & $\sim 0$ & $\sim 0$ \\ \hline
Incl. ND jets   & $\sim 0$/$\sim 0$ & 0.03/0.05 & 0.6/0.7  \\ \hline
\end{tabular}
\caption{\small{The same as in table~\ref{AllEY2.5} but for the enlarged tracker
  coverage $|\eta|<4.0$.}}
\label{AllEY4}
\end{center}
\end{table}
From the point of view of the phenomenological analysis presented here, the
situation may be improved by cutting on the variable proposed
in~\cite{HarlandLang:2011ih}, namely the maximum kinematically allowed values of
$m_{\tilde{\chi}}$ and $m_{\tilde{l}}$ assuming the signal decay chain.
%While many of the cuts in this paper (for example, on $W_{\rm miss}$) will be highly correlated with these, it is nonetheless possible that by using these variables, a more favourable signal to background ratio may be achieved. 
Following the approach of ~\cite{HarlandLang:2011ih}, we have checked that after
applying the FPD acceptance and lepton \pt\ and $\eta$ cuts, and requiring the
maximum neutralino mass to be larger than 100~GeV (to be consistent with the
$W_{\rm miss}>200$~GeV cut in table~\ref{cuts}), the signal yields are increased
by 50-80\% depending on the mass configuration, while the WW background remains
the same. We may also expect some reduction in the low mass dilepton SD and DD
backgrounds, but in the absence of a full MC implementation this cannot
currently be calculated.

Experimentally, the signal yield can be  doubled by taking all di--lepton masses
into account. This would, however, not only increase the background but also the
average di--lepton mass itself and hence limit the possibility of estimating the
unknown mass of the DM particle by measuring the central system mass via the
FPDs. Another way to increase the signal yield would be to increase the lepton
reconstruction efficiencies.
%In the ATLAS study \cite{Aaboud:2017leg} they start
%at 70\% for muons and at 50\% for electrons and slowly rise with increasing
%\pt, nevertheless since we deal with two leptons, any increase in the 
%single-lepton efficiency could have a reasonable impact.
The background contamination, in turn, could be lowered by rejecting events with
a displaced vertex which can be done by restricting track longitudinal, $z_0$,
and transverse, $d_0$, impact parameters to some small values, or by a cut on
the so called pseudo-proper lifetime. Furthermore, as discussed above, both
ATLAS and CMS are upgrading their trackers to cover the additional region
$2.4 < |\eta| < 4$. Both are also considering adding timing detectors in these
forward areas with resolution of about 30~ps \cite{CMSMIP,ATLASHGTD}.
By getting this timing information we acquire another ToF rejection factor in
addition to that shown in table~6 of Ref.~\cite{Harland-Lang:2018hmi}.

\section{Conclusions}\vspace*{-0.1cm}
We have discussed the prospects for searching for slepton pair production via
leptonic decays in compressed mass scenarios at the LHC, via photon--initiated
production. In this case the experimental signal is simple, comprising only four
charged particles in the final state, namely two forward outgoing protons and
two leptons in the central detector.

After accounting for all relevant effects, we find that the backgrounds from
pile--up and semi--exclusive photon--initiated lepton pair production are
expected to be of the same order as the signal, with the irreducible
$WW$ background being somewhat lower. 
We have also discussed a variety of ways in which this situation could be
improved upon, with the potential for increased tracker acceptance combined with
timing detectors at the HL--LHC being particularly promising. While a detailed
study of the possibilities at the HL--LHC is beyond the scope of the current
work, this provides a strong motivation for further work on this area, and for
collecting data with tagged protons there. Certainly the main backgrounds are in
principle reducible and therefore with further investigation we may be able to
reduce these further.
\vspace*{-0.1cm}
\acknowledgments{\vspace*{-0.1cm}
This work was supported by the Ministry of
Education, Youth and Sports of the Czech Republic under the project LG15052.
MGR thanks the IPPP at the University of Durham for hospitality.
VAK acknowledges the support from the Royal Society of Edinburgh Auber award.}

\vspace*{-0.1cm}
\bibliography{eps}{}
\bibliographystyle{JHEP}

\end{document}